# Sampling from the Solution Space and Metabolic Environments of Genome-Scale Metabolic Models

Haris Zafeiropoulos[1], Daniel Rios Garza[2]

[1] Department of Microbiology, Immunology and Transplantation, Rega Institute for Medical Research, Laboratory of Molecular Bacteriology, KU Leuven, 3000, Leuven, Belgium, haris.zafeiropoulos@kuleuven.be
ORCID: https://orcid.org/0000-0002-4405-6802
[2] Université Paris-Saclay, Institut national de recherche pour l'agriculture, l'alimentation et l'environnement (INRAE), PRocédés biOtechnologiques au Service de l'Environnement (PROSE), Antony, France, danielriosgarza@gmail.com
ORCID: https://orcid.org/0000-0003-3865-2146

## Abstract

Flux sampling is an analysis that, based on a distribution, picks randomly an efficient number of points from the solution space of a metabolic model. Unlike most constraint-based analyses, flux sampling does not require an objective function to optimize, allowing for the exploration of the whole spectrum of the phenotypes a species can exhibit. However, sampling can also be restricted to a subspace where a chosen objective reaches at least a specified fraction of its optimum. This targeted approach adds value when investigating phenotypes that are optimal for a specific function. Contrary to Flux Balance Analysis, which returns a single solution, sampling leverages statistical power to uncover phenotypes that otherwise would be masked. This can be especially useful when changing the conditions (medium) in which a species lives. Here, we highlight some state-of-the-art methods for applying flux sampling at Genome-Scale Metabolic Models in different scenarios, and we showcase flux sampling applications.

## Introduction

Flux sampling has proven itself rather useful [1] among the more than 100 methods available for constraint-based analysis (CBA) of a metabolic network [2].

From the stoichiometry of the reactions present in a metabolic model — a mathematical representation of cellular metabolism— the stoichiometric matrix $S$ derives a $m*n$ matrix where $m$ is the number of metabolites in the model, and $n$ the number of reactions. At the same time, its values are the stoichiometric coefficient of each metabolite that participates in a reaction. We call Genome-scale Metabolic Models (GEMs) models in which the complete set of metabolic reactions encoded by an organism's genome is accounted for. Since there are

$n$ reactions, each getting a flux value, we can represent them with the **flux vector** $v$ where $v_i$ is the flux of reaction $i$. Thus, the **net production and consumption** corresponding to a given flux vector can be expressed by the matrix product $Sv$, which sums up all the production and consumption rates for each metabolite across all reactions. Therefore, $Sv = dx/dt$ where $x$ is the **concentration vector** of the $m$ metabolites present in the model, also expresses the time derivative of the metabolite concentrations. Assuming a steady state, i.e., the production rate of each metabolite equals its consumption rate, we can express this **mass balance** as $Sv = 0$, where $v$ is the **flux vector**, i.e., a vector whose elements are the flux value of a reaction. The steady state assumption, when expressed as a set of equations, sets the original **constraints** of this space [3].

Once **reaction directionality** is incorporated—based on thermodynamic constraints, heuristics, or empirical evidence—we impose inequality constraints on specific fluxes of the form $v_i \geq 0$ for irreversible reactions. This gives rise to a **flux cone**: a convex, unbounded subset of flux vectors that satisfy the steady-state condition $Sv = 0$ and the directionality constraints (Fig.1A).

Additionally, **exchange reactions** are used to model the uptake and secretion of metabolites between the metabolic system and the environment. At the same time, **experimental data** (e.g., from fluxomics or enzyme capacity assays) can further constrain internal reactions. In both these cases, bounds are assigned like: $v_i^{\min} \leq v_i \leq v_i^{\max}$. These bounds define an $n$-dimensional box (or hyperrectangle), where $n$ is the number of reactions. By intersecting this box with the null space of the stoichiometric matrix $S$ (i.e., the set of flux vectors satisfying the steady state condition), we obtain a **convex polytope** known as the **flux space** or **solution space**, denoted:
$$\mathcal{F}_v = \left\{ v \in \mathbb{R}^n \mid Sv = 0, \ v_i^{\min} \leq v_i \leq v_i^{\max} \right\}$$

**Table 1:** Spaces and their corresponding geometric objects derived from a metabolic model and its constraints.

| Space | Constraints | Geometric Object |
|---|---|---|
| **Initial flux space** | $Sv = 0$ | Linear subspace (null space of $S$) |
| **Flux cone** | $Sv = 0$, $v_i \geq 0$ for all irreversible $i$ | Convex polyhedral cone |
| **Feasible flux space** | $Sv = 0$, $v_i^{min} \leq v_i \leq v_i^{max}$ | Convex polytope |
| **Thermodynamically feasible flux space** | $Sv = 0$, $v_{imin} \leq v_i \leq v_{imax}$, if $v_i \neq 0$, then $v_i G_i \leq 0$ | Non-convex polytope |

Constraint-based analysis methods are either biased or unbiased, based on whether they depend on the presence of an **objective function** or not. The objective function is a mathematical expression that quantifies a specific property of the system, which the model seeks to **optimize** [3]. The most typical kind of analysis in metabolic modeling, Flux Balance Analysis (FBA), solves such an optimization problem [4] using a pseudo-reaction that tries to represent the species' requirements to grow, defined as the production of biomass per unit of dry weight, e.g., one mmol of biomass per gram of dry weight, called **biomass function**.

Even if it is widely used, classic **FBA** suffers from several challenges, with two of the most notable being:
- Biology-oriented: a cell does not continually optimise a particular process, e.g., growth; see *"Sampling suboptimal scenarios"* chapter
- Topology-oriented: the flux vector supporting the objective value is not unique even if the optimal value of the objective function is unique.

Sampling has the advantage of being unbiased, meaning it does not require an objective function. As we will see later, sampling with an objective function is also possible, particularly when one is focused on quantifying fluxes under a specific environment, instead of enumerating global features and/or phenotypes of the model. This is the rationale underlying methods that search for Elementary Flux Modes [5] and Minimal Pathways [6].

Sampling aims to generate a representative set of flux vectors from the interior of the feasible flux space (i.e., the flux polytope; Fig. 1B, see Table 1), typically according to a specific distribution—most commonly, a uniform distribution. Other distributions have also been suggested, such as the **Boltzmann (or Gibbs) distribution** that **maximizes entropy** under certain constraints, often including a **fixed or desired average growth rate** [7]. Here we describe how to **uniformly** sample the space of fluxes under different scenarios, in a biased or an unbiased way, depending on the question the researcher aims to address and at which level—at the single-species or the community level.

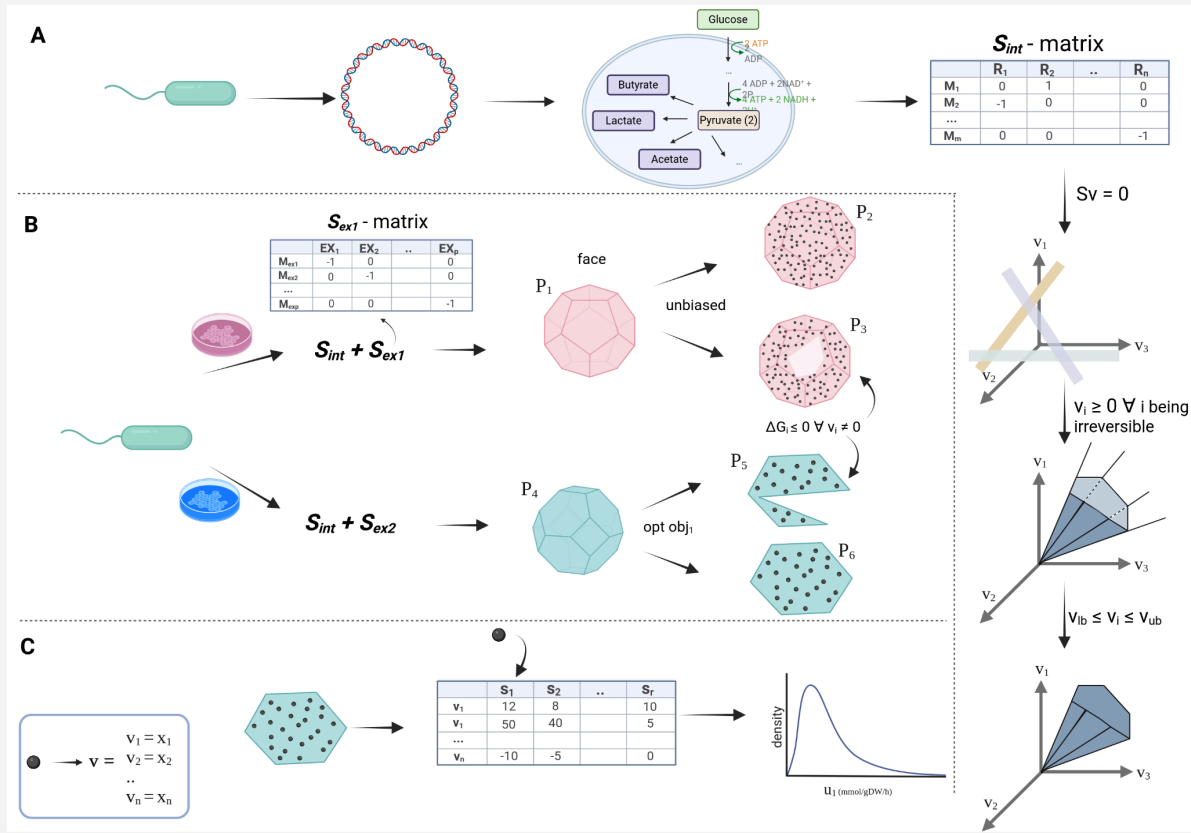

**Fig. 1:** Polytope construction representing environmental changes and optimization objectives, and flux sampling applications. **A.** From the genome of a strain, we can easily get the total set of enzymes it can express. Each enzyme catalyzes a specific reaction with fixed stoichiometry. Under the assumption of steady-state (i.e., no net accumulation of metabolites), the system of mass balance equations defines a flux space. Imposing reaction irreversibility constraints restricts this space to a **flux cone**. Further introducing lower and upper bounds on reaction rates transforms this cone into a **convex polytope**—the geometric object over which methods like flux balance analysis (FBA) and flux sampling operate. **B.** The environment a species grows in (medium) is also part of the model, with further reactions (called exchange reactions) providing what is available *in silico*, followed by their own stoichiometry and constraints. Therefore, different media lead to different polytopes ($P_1$, $P_4$). Flux sampling generates a sufficient number of points from the interior of the flux polytope, with each point being drawn according to a specified probability distribution. For example, under a **uniform distribution**, all feasible flux vectors within the polytope have an equal probability of being selected. One can sample either unbiased, with no objective function used, $P_2$, or in a biased way $P_6$. Considering that thermodynamics can benefit sampling by removing infeasible loops, but it turns the polytope into a non-convex polytope. Loopless sampling can be achieved either with special algorithms or by applying post-sampling methods. **C.** A sample consists of a flux value for each reaction in the model. From a sampling collection, one can extract the flux values corresponding to a specific reaction ($v_1$) and construct its marginal distribution; the probability of a single event occurring, independent of other events.

By sampling on the feasible solution space of the model, where all constraints are met, but none is getting optimal, one may get a comprehensive view of the solution space of the metabolic model, i.e., getting flux vectors that correspond to cases where biomass is or is not supported (assuming the biomass function does represent actual growth), or getting suboptimal values. Each sample corresponds to a unique flux vector that respects the model's constraints and thus, from a computational point of view, could be a "phenotype" of our *in silico* species (Fig. 1C). This way, we can reveal alternative flux states that may be biologically relevant but would be missed under FBA's assumption of a fixed optimal solution. This makes sampling a promising method for the study of metabolic shifts, both under the same and under different environmental conditions. In the first case, one would have to figure out whether there are clusters of samples within the same sampling experiment (dataset), while in the latter, across sampling datasets, corresponding to models with different media.

While valuable, flux sampling remains computationally intensive—particularly when aiming to capture the full range of possible growth scenarios for a species. To mitigate this cost, recent efforts have integrated flux sampling with topological approaches such as **Elementary Flux Modes (EFMs)** and **Minimal Pathways (MPs)**. MPs are defined as minimal sets of reactions within a subnetwork that must carry non-zero flux to satisfy all constraints imposed on the whole network. These approaches support arbitrary subnetworks and constraint configurations, offering a more scalable alternative for exploring feasible metabolic behaviors [6].

In addition, we will also discuss how one can sample the *in silico* environment of a species, as well as how to gain information about a community by sampling environments that support the steady-state biomass composition on a multi-species (community) level.

In the following section, we demonstrate how to efficiently apply uniform flux sampling on a series of scenarios. We highlight state-of-the-art sampling methods and key benchmarking results. Finally, we showcase the types of analyses that can be performed using the sampled flux distributions.

## Sampling scenarios

### Computational environment

All code to be shown in the following sections is online and can be found on the sampling branch of our metabolic_toy_model GitHub repository[1]. To perform the following methods, the user can either build a **GitHub codespace** or create a local conda environment as described in the Setting working space section of the flux_sampling.ipynb file; the basic notebook for this chapter. The dingo-based parts have been tested in Linux and MacOS (x86_64); they will soon be available for MacOS (arm64) and Windows.

1) Create a virtual environment, for example, using conda:

```
conda create -n sampling -y python=3.10
conda activate sampling
```

Or using venv on Windows:

---

[1] https://github.com/hariszaf/metabolic_toy_model

```
conda create -n sampling -y python=3.10
conda activate sampling
```

2) Clone the GitHub repository and install the dependencies:

```
git clone --branch sampling https://github.com/hariszaf/metabolic_toy_model.git
cd metabolic_toy_model
pip install -r requirements.txt
```

Solvers that will be discussed and tested to demonstrate their power are not actually required. However, one can get Gurobi and its corresponding license as described in a previous notebook of ours[2]. Concerning HIGHS [8], it is included as an extra dependency of pyoptinterface [9], which is specified and comes along with the rest of the dependencies of the dingo library.

## Random walks, solvers, and diagnostics

Flux sampling can be performed using a variety of algorithms, with Markov Chain Monte Carlo (MCMC) methods being widely used to address the inherent challenges of sampling high-dimensional spaces [10]. However, different geometric random walk strategies exhibit significantly different mixing times, affecting their convergence to the target distribution. We have previously demonstrated that, until recently and to the best of our knowledge, our Multiphase Monte Carlo Sampling (MMCS) algorithm and our implementation of the billiard walk remain among the fastest available methods. In the sampling scenarios discussed here, we rely on the dingo Python library [11], specifically leveraging its implementations of MMCS and the billiard walk. We also discuss the widely used Coordinate Hit-and-Run (CDHR) [12], as implemented in dingo, but also the Artificial Centering Hit-and-Run [13] and OptGp [14] samplers, since they are the two available on cobrapy; a core Python library for metabolic modeling.

Flux sampling is computationally expensive, and besides the sampling algorithms, the efficiency of the *solver* used—specialized software for solving linear programs—plays a critical role in overall performance. There is a great range of solvers, both open source and commercial, while *wrappers* like *PyOptInterface* support APIs for defining optimization problems in a solver-agnostic way, allowing a variety of underlying solvers to process them without changing the problem definition.

In this and the following sections, we use a GEM of the species *Blautia hydrogenotrophica* (Bh) as our case study. This species is an anaerobic species isolated from the human gut, known to be an acetogen containing the Wood–Ljungdahl pathway. We provide examples of how to sample using state-of-the-art approaches for flux sampling. First, we will use the methods supported by cobrapy.

---

[2] https://github.com/hariszaf/metabolic_toy_model/blob/duth/prep_env.ipynb

```python
import cobra
from pathlib import Path
root_dir          = Path().resolve()
agora_models_path = root_dir / "files" / "models" / "AGORA"
bh_agora_filename = "Blautia_hydrogenotrophica_DSM_10507.xml"

# Load model with cobra and set Gurobi as its solver
bh_cmodel = cobra.io.read_sbml_model(agora_models_path / bh_agora_filename)
bh_cmodel.solver = "gurobi"
```

Once the model is loaded, we can specify the solver of our choice. By default, this is typically GLPK, which can solve FBA problems for polytopes of any dimension, but struggles with high-dimensional cases—generally beyond a few dozen dimensions. Let us solve an FBA for our Bh model, to identify the optimal value that the objective function (by default, biomass) can take.

```python
bh_cmodel.optimize().objective_value
85.19
```

For more on how to interpret an FBA see Chapter XX.

Let us now perform unbiased sampling, meaning we are not enforcing any optimization objective. Instead, we simply seek flux vectors that satisfy the model's constraints, which in our case are limited to stoichiometric and directionality-based constraints. We first apply the OptGp sampler as implemented in cobrapy.

```python
# Unbiased sampling with OptGp as implemented in cobra;
optgp_sampler = OptGPSampler(bh_cmodel, thinning=100, processes=4)
optgp_samples = optgp_sampler.sample(n=10000)
```

Similarly, we can do the same using the ACHR sampler:

```python
# Unbiased sampling with ACHR as implemented in cobra;
achr_sampler = ACHRSampler(bh_cmodel, thinning=100)
achr_samples = achr_sampler.sample(n=10000)
```

In both cases, the interface looks identical; we select the sampler and, besides the model, we set two parameters, the number of samples to return (n) and the number of intermediate steps to be skipped between the recorded samples (thinning). Thinning has been found essential for the quality of our sampling [15, 16]; by sub-sampling the total chain, thinning allows us to reduce the autocorrelation among the generated samples. This increases the likelihood that the final samples are statistically independent—the higher the thinning, the better. OptGP was faster than ACHR (Table 2) for the same number of samples and the same thinning. The OptGP implementation in cobrapy can run in parallel (processes), making it better suited for larger models.

We can save our samples as pickle files, with a simple function like this:

```python
def dump_samples(samples, filename):
    # samples can be either a numpy.array or a pandas.DataFrame
    import pickle
    with open(filename, "wb") as f:
        pickle.dump(samples, f)

samples_path = root_dir / "results" / "samples"
dump_samples(achr_samples, samples_path / "achr_bh_unbiased.pkl")
```

Let us now sample, in an unbiased way, using dingo. To do so, we have first to silence the objective function of the model. By default, dingo considers whether an objective function is on, so to achieve unbiased sampling, we need first to turn off the objective.

```python
import dingo_walk as dingo

# Load Bh model as a dingo model
bh_posix  = (agora_models_path / bh_agora_filename).as_posix()
bh_dmodel = dingo.MetabolicNetwork.from_sbml(bh_posix)

# Replace the objective function with an empty zero-array
obj_fun = np.zeros(bh_dmodel.num_of_reactions())
bh_dmodel.objective_function = obj_fun
```

Now, we can perform sampling using the MMCS algorithm:

```python
# Sample using MMCS
mmcs_sampler = dingo.PolytopeSampler(bh_dmodel)
mmcs_samples = mmcs_sampler.generate_steady_states(ess=1000, psrf=True)
```

The last command will log in your monitor something like the following:

```
phase 1: number of correlated samples = 3600, effective sample size = 25, ratio of the maximum singilar value over the minimum singular value = 22012.3
phase 2: number of correlated samples = 3300, effective sample size = 1077

[1]maximum marginal PSRF: 2.72139
[2]maximum marginal PSRF: 1.00518
[2]total ess: 1077
```

Likewise, we can sample using any of the wide range of walks supported by dingo; among those cdhr and billiard_walk:

```python
# Sample with billiard walk as many samples as in the ACHR & OptGP cases
# It is always a good practice to load again your model, after you have used it
# to solve a set of linear programs
bw_sampler = dingo.PolytopeSampler(bh_dmodel)  # bh_dmodel is a fresh instance

bw_samples = bw_sampler.generate_steady_states_no_multiphase(
    method   = "billiard_walk",
    n        = 10000,
    burn_in  = 10,
    thinning = 100
)
```

Interestingly, in all cases of unbiased sampling on the Bh model, the marginal distribution of the biomass flux has its main mass close to 8 (Fig. 2). However, the distributions are not always the same, with OptGp returning a normal one, while BW produced a narrower one, with more samples grouped closely around the mean.

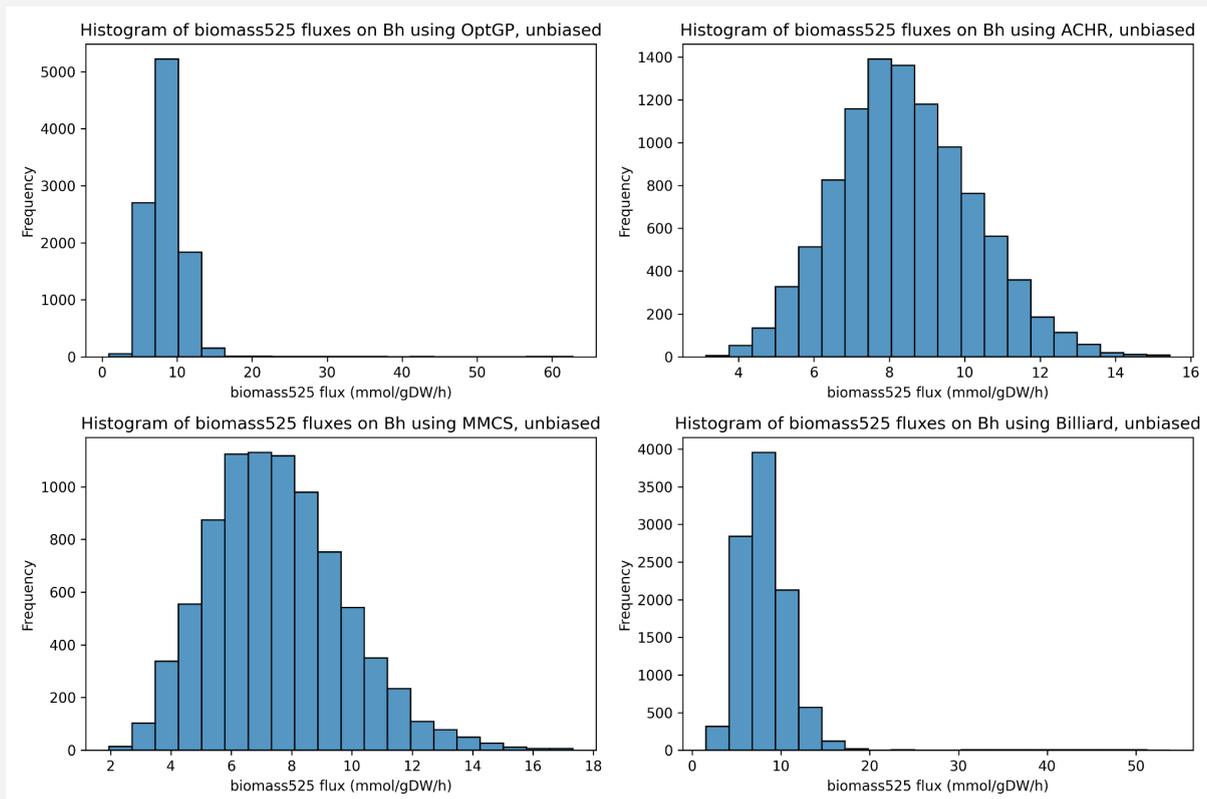

**Fig. 2:** Marginal distributions of the biomass flux after unbiased sampling on the solution space of Bh, using the OptGP (up-left) and the ACHR (up-right) samplers, as implemented in cobra, and the MMCS (down-left) and the Billiard walk, as in dingo (down-right). The optimal solution for biomass on Bh using this medium, was found to be equal to 85.19. Using the D'Agostino and Pearson's test for normality, the OptGP was the only case where the distribution was found normal. In all cases, the great mass of the distribution is close to 7.5. Billiard is the one that manages to sample states with a biomass flux higher than 50.

In the MMCS case, the parameters to be set are the convergence diagnostics Effective Sample Size (ess) and potential scale reduction factor (psrf). The ESS indicates how many effectively independent samples the correlated samples are equivalent to, while the PSRF assesses whether the chains have converged to the target distribution (in our case, the uniform distribution). When applied, after each phase of the MMCS algorithm, dingo computes the PSRF for each univariate marginal of the samples produced, sorts them, and checks whether the maximum is lower than 1.1, a commonly used threshold [17].

Now we can go through the message in the console we got when running the MMCS case. During the first phase, 3600 samples were produced in total, corresponding to 25 independent ones (ESS), while the maximum PSRF recorded for a reaction's marginal flux distribution was 2.7. In the second phase, 3300 samples were produced in total, being equivalent to 1077, and the maximum PSRF was 1.005; since we had asked for a ess of 1000 the process is complete. This experiment is indicative of how effective the first phase of MMCS is.

To have an estimate of the ESS of the other methods, we can use the arviz Python library:

```python
import arviz as az

# Make sure your dataframe has samples as rows and reactions as columns
df = load_samples(samples_path / "achr_bh_unbiased.pkl")
data = df.values
# If you are aware of the number of chains produced during sampling,
# replace None with the number of chains and edit the shape accordingly.
# For example, if you had 10000 samples coming from 4 chains: [4, 2500, :]
data    = data[None, :, :]
dataset = az.convert_to_dataset({"samples": data})
# Estimate ess
ess = az.ess(dataset)
```

You can then check the ess.mean() and the ess.min() to have an estimate of your sampling. A low min means that some reaction has a very low ESS. You can identify which reactions have a low ESS.

```python
ess_values        = ess['samples'].values
low_ess_idx       = (ess_values < 1000).nonzero()[0]
low_ess_reactions = df.columns[low_ess_idx]
print(low_ess_reactions.values)
```

Table 2 shows some total converging times for the walks and the solvers discussed; since MMCS does not have number of samples as an argument, we calculated the estimated ESS of the rest of the experiments and kept the highest (~3000) to provide it as an argument on MMCS, while PSRF was also set to true.

**Table 2:** Computing time required for sampling in an unbiased way on the original flux space of the Bh model, using the two available sampler algorithms on cobra, the MMCS algorithm and the Billiard walk as implemented in dingo.

| Sampler | Solver (Default/Gurobi) | Time (s) | Notes |
|---|---|---|---|
| OptGP | GLPK (default) | 90.48 | With 4 processes |
| | Gurobi | 80.42 | |
| ACHR | GLPK (default) | 235.16 | |
| | Gurobi | 181.53 | |
| MMCS | HiGHS (default) | 56.16 | PSRF set to True and ESS equal to highest of the 6 other sample datasets |
| | Gurobi | 54.55 | |
| Billiard | HiGHS (default) | 168.88 | |
| | Gurobi | 174.58 | |

Last, you may notice that in the dingo runs, we did not set a specific solver. By default, dingo uses HIGHS. We can change the solver by:

```
mmcs_sampler.set_solver("gurobi")
mmcs_sampler._parameters
{'tol': 1e-06,
 'solver': 'gurobi',
 'distribution': 'uniform',
 'nullspace_method': 'sparseQR',
 'opt_percentage': 100,
 'first_run_of_mmcs': False,
 'remove_redundant_facets': True}
```

Gurobi and HIGHS required almost the same time in dingo; in the case of the Billiard walk, HIGHS was even faster. The default algorithm in cobra though, GLPK, required much longer than Gurobi both for OptGP and ACHR. This is i

## Rounding, sampling, and back-transform

In the previous paragraph, we showed how different samplers and solvers may affect both the quality and the efficiency of our sampling. Choosing the optimal walk or configuration is not straightforward, and it is unlikely that a single sampler will perform best in all scenarios. In addition, there are cases where even well-established software fails to sample on the flux space of a metabolic model, not only because of its high dimensionality, but also because of its topology. This is because polytopes derived from metabolic models tend to be highly anisotropic, or "skinny." To address this, a common strategy involves an initial rounding step: the polytope is transformed into a more isotropic shape to facilitate more efficient sampling, and the resulting samples are then mapped back to the original space.

A state-of-the-art software for this task is *PolyRound* [18]. PolyRound removes redundant constraints, refunctions inequality constraints that are de facto equalities, and embeds the polytope in a space where it has non-zero volume. In this *"rounded"* polytope, sampling is much easier to achieve. In the following, we show how one can round the initial polytope using *PolyRound* and perform sampling using the MMCS and the Billiard walk as in *dingo*.

```python
from PolyRound.api import PolyRoundApi
from PolyRound.settings import PolyRoundSettings
from PolyRound.static_classes.lp_utils import ChebyshevFinder

# Fix settings
settings          = PolyRoundSettings()
settings.verbose = True
settings.backend = "gurobi"

# Export initial and perform rounding tasks
bh_modelfile          = agora_models_path / bh_agora_filename
polytope              = PolyRoundApi.sbml_to_polytope(bh_modelfile)
polyrounded_polytope = PolyRoundApi.simplify_transform_and_round(polytope)
```

Then you are ready to sample using the polyround-ed polytope; here is how you would do this with dingo and any walk it supports:

```python
# Sample on polyrounded polytope using MMCS
mmcs_poly_samples = dingo.PolytopeSampler.sample_from_polytope(
    np.asarray(polyrounded_polytope.A),
    np.asarray(polyrounded_polytope.b),
    ess  = 1000,
    psrf = True
)

# Essential step! After sampling, we need to map our samples, back to the
original polytope
mmcs_poly_samples = polyrounded_polytope.back_transform(mmcs_poly_samples)
```

As shown in the chunk above, after sampling, we again use the polyround-ed polytope to apply a back-transformation. This is crucial since our samples now correspond to a lower-dimensional space, i.e., they do not correspond to a single reaction. In the following chunk, one can see the number of reactions of the Bh model we use, and the dimensions of the samples before and after the back-transformation.

```
>>> len(bh_dmodel.reactions)
1068
>>> mmcs_polyrouned_samples.shape
(179, 5500)
>>> mmcs_poly_samples.shape
(1068, 5500)
```

Likewise, we can use the polyround-ed polytope with non-MMCS walk on dingo. For example, to sample with the Billiard walk:

```
bw_polyrounded_samples = dingo.PolytopeSampler.sample_from_polytope_no_multiphase(
    np.asarray(polyrounded_polytope.A),
    np.asarray(polyrounded_polytope.b),
    method   = "billiard_walk",
    n        = 10000,
    thinning = 100
)
# Like, in the MMCS case, remember to apply the back transformation
bw_poly_samples = polyrounded_polytope.back_transform(bw_polyrounded_samples)
```

In the same notion, any sampling algorithm can be applied.

## Computational vs. biological feasibility

In this section, we highlight key biology-related but modeling-oriented challenges to consider when applying flux sampling. We also present examples of basic post-sampling analyses that can be used to extract insights from the generated samples.

In the previous section we sampled in an unbiased way, using a medium which from now on we will call *rich*; one in which every exchange reaction in the model has a non-zero value in the medium. Here is how you can confirm that our original model has a *rich* medium:

```
# Check whether there are exchange reactions on the model, not present in the medium
missing_exchanges = [
    r.id for r in bh_model.exchanges if r.id not in bh_model.medium
]
print(missing_exchanges)
# Check whether there is any zero flux case in the medium
print(np.count_nonzero(list(bh_model.medium.values())))
111
```

However, one can easily notice a rather problematic case:

```
# Load dingo samples from unbiased sampling with MMCS
unb_samples = load_samples(samples_path / "gurobi_MMCS_unb.pkl")
# Build a df with the model reactions as column names
unb_samples = pd.DataFrame(unb_samples, index = bh_reactions).T
# Get mean value of the oxygen uptake among the samples
unb_samples["EX_o2(e)"].mean()
-621.8355798043938
```

This is not a sampling issue. If you try an FBA, even a ***parsimonious*** one, on top of the FBA optimization it also minimizes the total sum of flux, or a ***loopless*** one, where thermodynamic infeasible inner cycles are removed, you will still get a similar result where oxygen is being "used". However, it is well known that *Blautia* does not use oxygen, in most cases is not even tolerant to it. This brings us in front of a major challenge in metabolic modeling: mass balance and directionality constraints alone, most often than not, are enough to sufficiently constrain the total solution space to a biologically-feasible one. In this case, we observe a case where even when thermodynamics are applied is not enough. This is because of the reactions present in the model: one can easily observe that when oxygen is present, the optimal growth of the model is higher compared to when is absent:

```
# Objective value for biomass using the original, rich medium
bh_model = cobra.io.read_sbml_model(bh_xml)
init_sol = bh_model.optimize()
85.1972174193773
# Objective value for biomass after removing oxygen from the medium
bh_model.reactions.get_by_id("EX_o2(e)").lower_bound = 0
sol_no_o2 = bh_model.optimize()
sol_no_o2.objective_value
78.25516328464289
```

This highlights the need for further curation of the models [19]. Yet this goes beyond the scope of this chapter. Given this drawback, we can still use this GEM to qualitatively study a series of questions, always keeping in mind that if we are interested in more accurate predictions, extra work on the model itself is required. For example, in our unbiased sampling data-set, how different are the samples returned? Do they build groups? Are there coupled reactions and can we infer pathways working together or canceling on the other?

## Basic statistics on sampling data sets

To address questions like those in the end of the previous section, one can apply well established statistical techniques. For example, one can have an overview of how different the samples are by performing a simple Principal Component Analysis (PCA). By examining the first principal component, one can identify which reactions contribute most significantly to the differences observed across the samples.

```
from sklearn.decomposition import PCA

def pca_samples(samples):
    # Run a PCA for 2 principal components
    model = PCA(n_components=2).fit(samples)
    n_pcs = model.components_.shape[0]
    # Get reactions with highest contribution in each component
    most_important = [
        np.abs(model.components_[i]).argmax() for i in range(n_pcs)
    ]
    most_important_names = [
        samples.columns[most_important[i]] for i in range(n_pcs)
    ]
```

```python
    # Return them as a dictionary
    return {'PC{}'.format(i): most_important_names[i] for i in range(n_pcs)}
```

Using again the Bh model, we can remove oxygen from the medium by setting its lower bound (the one standing for the uptake flux bound) to zero, and sampled again using the MMCS algorithm and the Gurobi solver:

```python
# Build a new cobra instance
bh_model_no_oxygen = cobra.io.read_sbml_model(bh_xml)

# Set lower bound of the uptake reaction for oxygen to zero, "removing" it from
the medium
bh_model_no_oxygen.reactions.get_by_id("EX_o2(e)").lower_bound = 0

# Build a new dingo Model from this cobra model
dmodel = dingo.MetabolicNetwork.from_cobra_model(bh_model_no_oxygen)

# Replace its objective function with the zero-vector
dmodel.objective_function = np.zeros(len(dmodel.reactions))

# Sample unbiased using MMCS
sampler = dingo.PolytopeSampler(dmodel)
unb_anox_samples = sampler.generate_steady_states(ess=1000, psrf=True)
```

When we apply the pca_samples() function on our unbiased samples that no longer use oxygen, we observe that there are no patterns and the first two PCs only explain a very low diversity of the samples (Fig. 3). Moreover, we come across the limitations described in the previous section again; the reaction contributing the most to the first PC is EX_lac_D(e). It would make sense to check how this reaction splits the samples.

```
pca_samples(unb_anox_samples, plots=True)
Explained Variance Ratio: [0.06752154 0.04478603 0.04003647 0.03823201
0.03748612 0.03641162
 0.03585641 0.03405611 0.03327041 0.03257265]
{'PC0': 'EX_lac_D(e)', 'PC1': 'EX_acald(e)'}
```

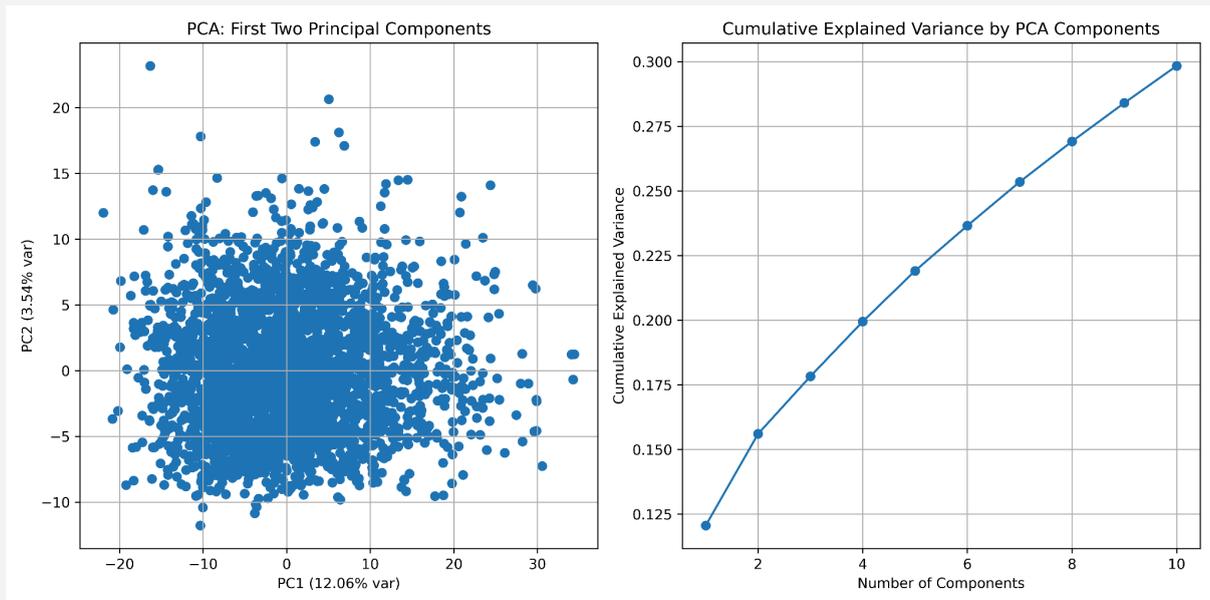

**Fig. 3:** PCA on unbiased sampling on Bh model after removing oxygen from its medium. The roughly uniform distribution of the first two principle components (left) indicates no apparent clusters or patterns among the samples. On the right, the first ten principal components cumulatively explain 30% of the sample variance, indicating there might be some structured variance, but it is spread across many components.

From its minimum and maximum values, it is clear that there are cases that lactate is being produced, while in others it is consumed:

```
print(
    unb_anox_samples["EX_lac_D(e)"].min(), unb_anox_samples["EX_lac_D(e)"].max()
)
-999.6086122899679 999.5103709847976
```

When one is interested in a specific reaction, such as lactate exchange, you can split your dataset into two groups: one with a positive flux for this reaction and another with a negative flux. You can then identify reactions that differ significantly between the two groups using a statistical test such as the Mann–Whitney U test, combined with a multiple-testing correction, e.g. false discovery rate, Bonferroni etc . From the resulting set of reactions, you can further narrow the list by retaining only those whose flux sign also changes between the two groups.

```
# Perform Mann-Whitney test after splitting samples in two groups:
# those producing lactated - those consuming it
(
  sig_cols, pos, neg, dddf
) = different_fluxes_over_sign(unb_anox_samples, "EX_lac_D(e)", bh_model)

# Reactions with a mean of a different sign between the two groups
dddf["Reaction name"]
0               D-Lactate Transport via Proton Symport
```

```
1                          L-Lactate exchange
2                       D-Lactate Dehydrogenase
3                       L-Lactate Dehydrogenase
4    L-Lactate Reversible Transport via Proton Symport
5           Nucleoside-Diphosphate Kinase (ATP:dCDP)
Name: Reaction name, dtype: object
```

However, both D- and L-Lactate Dehydrogenase are thermodynamically unfavorable when trying to convert lactate to pyruvate[3,4]. We will discuss below how to address the thermodynamics challenge, yet it is becoming clear how one can apply basic statistical tests on their samples. Dimensionality reduction and clustering techniques can be applied to identify potential groups or phenotypic states within the sample space. To facilitate such analyses, a new Python library called dingo-stats has been initiated[5], offering a foundational toolkit for these tasks.

## Biased sampling: optimal and suboptimal cases

So far, all our sampling experiments have been unbiased. Yet, sampling can be performed by specifying an objective function to be optimized, i.e. sampling on a face of the feasible solution polytope, or by sampling within a subspace defined by constraints that enforce a minimum (sub-optimal) objective value (Fig. 1).

**Suboptimality** in metabolic systems may offer increased robustness to stochastic variation and perturbations, enhancing the stability and functionality of biological networks [20]. Enzyme and metabolite levels fluctuate almost constantly in most biological systems. Thus, tolerance to such suboptimality allows metabolic networks to maintain functionality under varying conditions, fostering a more resilient system [21]. This is particularly evident in bacteria, where suboptimal resource allocation in fluctuating environments can constrain their growth and response, yet provide the flexibility necessary for adaptation [22]. In evolutionary contexts, mutants typically exhibit suboptimal flux distributions that lie between the wild-type optimum and the mutant's specific optimum, further illustrating the balance between optimal and suboptimal states in sustaining cellular function and adaptation [23].

We first sample the Bh model, removing oxygen from its medium, and forcing lactate uptake to be zero, those two are actually the same thing, asking for an optimal biomass; by default dingo is using the default objective that in all AGORA models is the biomass function. We use Billiard walk and Gurobi:

```
def model_prune_medium(exchanges_to_mute=[]):

    bh_model_no_oxyg = cobra.io.read_sbml_model(
        agora_models_path / bh_agora_filename
```

---

[3] https://tinyurl.com/23mhhxjn
[4] https://tinyurl.com/vvt5ds7h
[5] https://github.com/SotirisTouliopoulos/dingo-stats/

```python
    )
    for rxn_id in exchanges_to_mute:
        bh_model_no_oxyg.reactions.get_by_id(rxn_id).lower_bound = 0

    d_model = dingo.MetabolicNetwork.from_cobra_model(bh_model_no_oxyg)

    return d_model

d_model = model_prune_medium(["EX_o2(e)", "EX_lac_D(e)"])
sampler = dingo.PolytopeSampler(d_model)
sampler.set_solver("gurobi")
no_lac_ox_opt  = sampler.generate_steady_states(ess=2000, psrf=True)
```

To sample suboptimal cases, we follow the same steps but this time we use the set_optimal_percentage method of the dingo.MetabolicNetwork class:

```python
# Build a bh_model_no_o2_lac as in the chunk above, and then a dingo model based
on that
d_model  = dingo.MetabolicNetwork.from_cobra_model(bh_model_no_o2_lac)
# Set optimal percentage to 50, i.e. 50% of optimal value
model.set_opt_percentage(50)
sampler = dingo.PolytopeSampler(model)
# Now you can sample as above to get no_lac_ox_opt50
```

One can easily detect how absolute optimality affects flux variability by comparing the number of fixed fluxes in these two cases:

```python
from scripts.stats import check_samples_range
# Part of check_samples_range()
threshold = 0.001
non_fixed_fluxes = [
   col for col in samples if abs(samples[col].min() - samples[col].max()) >
threshold
]
check_samples_range(no_lac_ox_opt)
Number of reaction with a non-fixed flux: 155

check_samples_range(no_lac_ox_opt50)
Number of reaction with a non-fixed flux: 619
```

As in the unbiased case, we can apply statistical tests on the samples. For example, we may apply correlation tests for groups of reactions of interest, such as those forming a subsystem on our GEM. Since Bh is an acetogen, we focus on the subsystem associated with the acetate kinase (ACKr) reaction, i.e. pyruvate metabolism: (Fig. 4, left)

```python
from scripts.stats import correlated_reactions
```

```
# Get the subsystem that the acetate kinase is part of
subsys = bh_model.reactions.ACKr.subsystem
cor_matrix, cor_dict = correlated_reactions(
    no_lac_ox_opt_clip, rxns_per_subsystem[subsys],
    subsystem=subsys, outfile="opt"
)
```

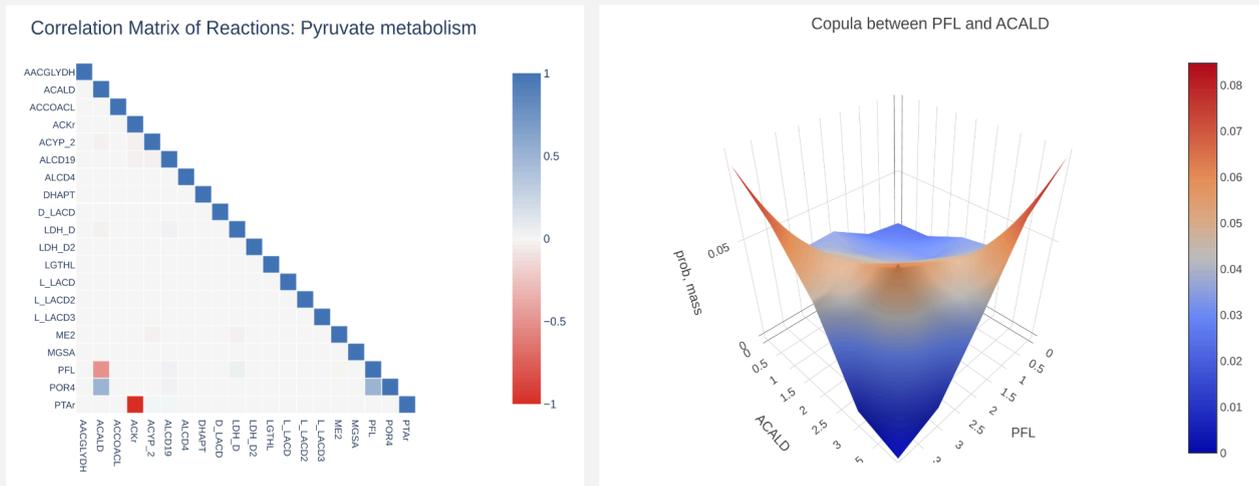

**Fig. 4:** Sampling Statistical Power with ≥50% optimal biomass flux in Bh after oxygen and lactate uptake shutdown. Correlation heatmap for the reactions that belong in the pyruvate metabolism subsystem (left). Bh is a known acetogen. ACKr (acetate kinase) which is the main reaction producing acetate, shows a strong negative correlation with PTAr (phosphotransacetylase), while ACALD (acetaldehyde dehydrogenase) is negatively correlated with PFL (pyruvate formate lyase) and positively correlated with POR4 (pyruvate-flavodoxin oxidoreductase). Using copulas one can visualize the dependency between two fluxes. In this case, the negative association between ACALD and PFL is shown, as ACALD attains its lowest values when PFL reaches its highest, and vice versa. Apparently, one can use any reaction set of interest, combine reactions from different subsystems or describe their own subsystems. In the same notion, one can check how the flux of a reaction depends on the one of another using copulas (Fig. 4, right).

```
dingo.illustrations.plot_copula(
    data_flux1 = [no_lac_ox_opt50_clip["ACALD"], "ACALD"],
    data_flux2 = [no_lac_ox_opt50_clip["PFL"], "PFL"],
    n = 5
)
```

Asking for biomass to get at least some minimum positive value, we set an essential constraint for our model that leads to biologically more accurate samples since ACKr gets only negative values now, while in the unbiased case would range free.

```
no_lac_ox_opt50["ACKr"].min(), no_lac_ox_opt50_clip["ACKr"].max()
(-999.7370756786753, -0.21382671605947887)
bh_unb_anoxic["ACKr"].min(), bh_unb_anoxic["ACKr"].max(),
(-996.283249394052, 317.50328690947606)
```

However, the need for further model curation and thermodynamics constraints remains essential for more accurate flux vectors.

## Thermodynamically feasible sampling

As shown in the previous sections, mass balance and directionality and inhomogeneous linear constraints, representing directionality, alone, most often than not are not enough to sufficiently constrain the total solution space to a biologically-feasible one. Lack of thermodynamic

Yet , flux sampling the solution space of GEMs considering thermodynamic constraints is not straightforward yet. Two main approaches for reaching thermodynamically feasible samples have been used:
- applying post-sampling steps to *"push"* a sample to its closest thermodynamically feasible one. It has been shown that the latter will not result in a uniform distribution over the thermodynamically-feasible portion of flux space.
- adding such constraints on the model and sample on a non-convex, non-connected polytope; the most well known algorithms and software designed for this purpose include: ll-ACHRB [24], LooplesssFluxSampler [25], Gapsplit [26], pta[27]

However, each of the software mentioned above comes with its own limitations and applying them in GEM is still not completely straightforward. For example, the Probabilistic Thermodynamic Analysis (PTA) that models the uncertainty of free energies and concentrations with a joint probability distribution, returns samples of high biological relevance. Yet, it requires model preprocessing as described in the supplementary material on the corresponding study, where the authors have used a set of *NetworkReducer* features and manual curation steps.

A much easier approach is to apply a loopless Flux Variability Analysis (FVA) for optimizing a biologically relevant objective reaction, e.g. the biomass, ATP production etc, and apply its findings as lower and upper bounds of your model. Then you can apply sampling and check whether your resulting samples are loopy or not. After you fire again a new model without lactate and oxygen, you may run:

```
opt50_no_ox_lac_ll = run_pseudo_ll_sampling(bh_model_no_o2_lac)
```

You can then apply statistical methods, such as correlation analysis, as shown above.

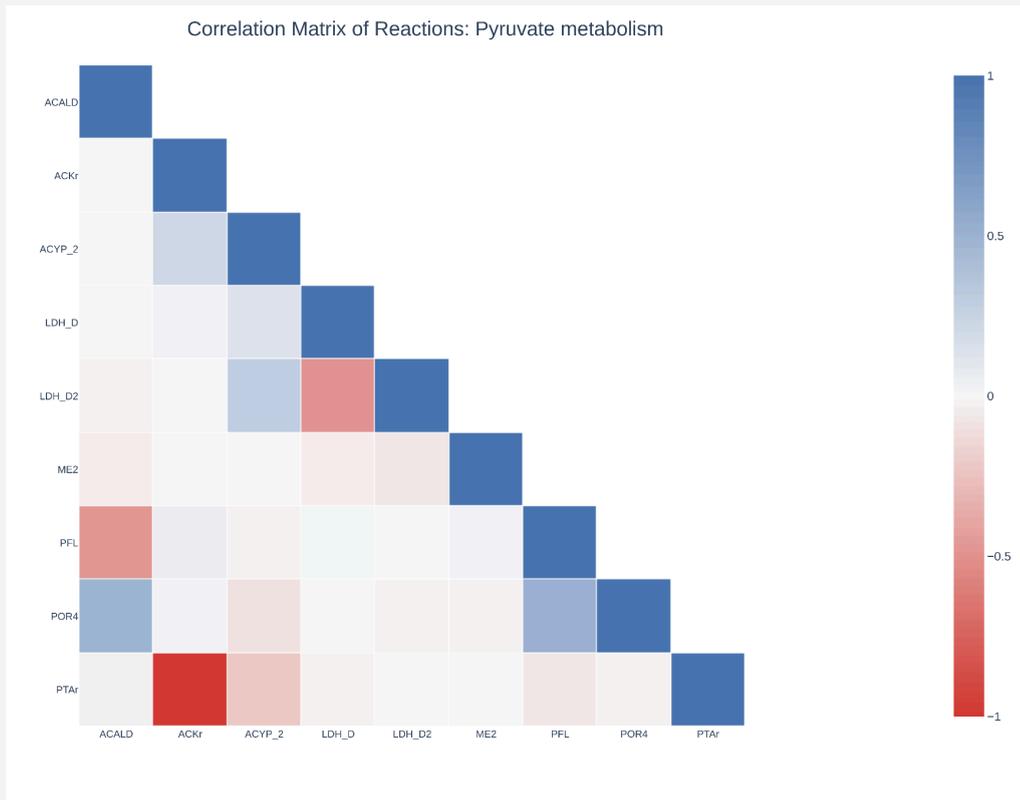

**Fig. 5:** Correlation matrix of the pyruvate reactions after applying loopless FVA–derived constraints to the Bh model, followed by sampling constrained to achieve at least 50% of the optimal biomass. You may notice that some reactions compared to Fig. 4 are missing, this happens since those get fixed values.

This way, sampling depends on the presence of an objective function, since FVA requires one. Also, even if in our sampling we are asking for a half-optimal biomass, when applying loopless FVA we had a fraction of optimum of exactly one; meaning that the fixed constraints as in the case of optimal growth, stay fixed even in our sampling. As shown in Fig. 5 though, the remaining reactions are much more correlated than before.

## Sampling the Metabolic Environment

We now briefly focus on exchange reactions—the interface between the model and its environment. Every cell is embedded in a metabolic environment, and the distribution of intracellular fluxes depends directly on access to external metabolites. Although extracellular concentrations or relative abundances of metabolites are not easily convertible into flux values, their supply and limitations influence the flux distribution. To capture this influence, the environment is typically modeled as a set of constraints—bounds on the rates at which metabolites can be imported or exported.

While this does not fully capture the quantitative complexity of real environments, sampling methods make it possible to at least qualitatively explore how flux distributions depend on environmental conditions. This allows us to distinguish between environment-driven reactions, whose activity is sensitive to environmental constraints (reflecting the relative availability of external resources), and environment-neutral reactions, which consistently carry flux regardless of those constraints.

In some of our projects, we simulate different environmental contexts by sampling nutrient availability from a Dirichlet distribution parameterized by a vector of ones. The Dirichlet is a natural choice for this task: it defines a uniform probability density over the simplex—the space of all possible relative compound compositions that sum to a fixed total. Geometrically, it acts like a high-dimensional ball over all admissible uptake combinations, ensuring that each sampled environment represents a distinct but unbiased composition.

This approach enables us to systematically explore how variation in nutrient proportions influences feasible metabolic states. More importantly, sampling broadly and without bias allows us to identify metabolic properties that remain robust across environmental compositions.

In this example, we will perform the following for the species *Bacteroides thetaiotaomicron*, sampling fifty metabolic environments. We will:

1) Retrieve the extracellular metabolites that our species potentially uses.
2) Generate a Dirichlet distribution, which we call an environment ball, whose sum is fixed to a value for the uptake of these extracellular metabolites.
3) Simulate our model on each of these environments, retrieving and, finally, comparing steady-state flux distributions across environments.

Functions to accomplish these tasks are provided in the script envBallScripts.py:

1) Import the environment ball utilities, load the model and extract exchange reactions:

```python
from scripts.envBallScripts import *

root_dir = Path(__file__).resolve().parent.parent
model_folder = root_dir / 'files' / 'models' / 'AGORA'

model = cobra.io.read_sbml_model(model_folder /
'Bacteroides_thetaiotaomicron_VPI_5482.xml')

exchanges = get_exchange_metabolites(model)
```

2) Generate 1,000 Dirichlet-sampled environments. The total influx is set to 100 mmol/gDW/h. We also set oxygen to zero and fix water bounds to 100.

```python
envBall = gen_environment_ball(
    exchanges,
    anaerobic       = True,
    fixed_reactions = {'EX_h2o(e)': 100},
    size            = 1000,
    total_flux      = 100,
```

```
    seed            = 666
)
```

One can fix additional metabolites (e.g. trace elements) by adding their exchange reaction ids and maximum flux bounds to the "fixed_reactions" dictionary.

3) Simulate our model under each environment and build a clustermap with environments as columns and reactions as rows:

```
solutions = apply_env_ball(model, envBall)

results_folder = root_dir / 'results' / 'env_ball'
os.makedirs(results_folder, exist_ok=True)

outputPath = results_folder / 'env_ball_reactions_cluster.png'

plot_flux_heatmap(solutions, outputPath)
```

## Sampling the pangenome (or panreactome)

All the methods described so far are based on individual metabolic reconstructions from sequenced genomes. However, microbiome studies often generate data at higher taxonomic levels than isolated strains. In such cases, depending on the scientific question of interest, it might be more relevant to sample from pan-genomes--i.e., the union of genes within a taxonomic clade (species, genus, family, class, or phylum). More specifically, we focus here on the **panreactome**: the union of metabolic reactions across a taxonomic clade.

In this section, we explain how to reconstruct a panreactome as an SBML object, using the same format as we used for isolated genomes. Once this object is created, we can apply the same sampling techniques described earlier. We will also use this object in the next section to reconstruct pan-elementary flux modes (pan-EFMs) that can reveal non-neutral gene essentiality at higher taxonomic levels [28].

In this example, we build a panreactome for the *Bacteroides* genus. The general procedure is as follows:
1) Place single-genome models of the target taxon into a folder.
2) Load each model, removing exchange reactions and the objective function.
3) Build a new model containing the set of reactions that occur at least one genome-scale model.
4) Add exchange reactions based on transporters.
5) Add the objective function and save the resulting SBML file.

This SBML file can be used for sampling the pan-reactome instead of sampling individual genome-scale models.

Functions to accomplish these tasks are provided in the script build_pan_reactomes.py.

1) Load the folder containing *Bacteroides* models and extract store a biomass objective function:

```
from build_pan_reactomes import *

model_folder = agora_models_path / 'pan_genome'
models       = os.listdir(model_folder)
bt_xml_file  = 'Bacteroides_thetaiotaomicron_VPI_5482.xml'
model = cobra.io.read_sbml_model(agora_models_path / bt_xml_file))

for reaction in model.reactions:
    if reaction.objective_coefficient==1.0:
        objective = reaction.copy()
```

2) Build the panreactome. Before adding the exchange reactions or the objective function, generate a heatmap showing the presence and absence of reactions across the *Bacteroides* genus. Reactions in the columns are sorted by frequency across models; Models in the rows are clustered using Hamming distance and average linkage.

```
preact = make_panReactome(model_folder, 'Bacteroides_panreactome')

# Where we save the image and reaction presence absence table
results_folder = root_dir / 'results' / 'pan_reactome'

generate_binary_presence_matrix(model_folder, preact, results_folder)
```

3) Add exchange reactions and the objective function. Optimize the panreactome model, print the solution, and save it as an SBML file in the results folder.

```
add_exchange(preact)
preact.add_reactions([objective])
preact.reactions.get_by_id(objective.id).objective_coefficient = 1.0

solution  = preact.optimize()
print(f"Objective Value: {solution.objective_value: .2f}")

cobra.io.write_sbml_model(
    preact, os.path.join(results_folder, 'bacteroides_pan_reactome.xml')
)
```

The expected output is:
```
Objective Value:  191.76
```

## Sampling PanEFMs

Next, we consider sampling the solution space of genome-scale metabolic models from a different perspective. What if, instead of drawing samples from the interior of the convex polytope, we are interested in understanding the structure of the space itself?

The sampling methods introduced so far produce uniform samples from the space of feasible metabolic flux distributions. However, an alternative approach to investigate the possible ways for metabolism to function is to study the distribution of reactions that necessarily function together—i.e., **elementary flux modes (EFMs)**.

By definition, EFMs are irreducible sets of reactions that can carry a steady-state flux. If any reaction in the set is removed, the remaining reactions can no longer support flux. How reactions are distributed across different EFMs reveals several important structural properties of genome-scale models and the space of metabolisms they encode.

Two illustrative cases:

1. If a reaction is present in all EFMs, then it is essential for the metabolic network.
2. If a reaction appears in all EFMs under one environmental condition, but not in others, then it is conditionally essential, required for survival in that specific environment.

To illustrate these concepts, we apply them to the panreactome constructed above. Specifically, we sample random elementary flux modes (EFMs) from the *Bacteroides* panreactome under different, also random, environmental conditions (see sampling from the environment). Based on this sampling, we classify reactions into three categories:

1. Super-essential reactions – reactions that are required in all sampled EFMs across all environments; these represent core functions necessary for any feasible *Bacteroides* metabolism.
2. Environment-specific reactions – reactions that are essential only under particular environmental conditions, indicating environment-specific metabolic requirements.
3. Dispensable reactions – reactions that are rarely or never needed in the sampled EFMs, and thus represent metabolic functions that are either redundant, neutral, or result from data artifacts (such as belonging to not well-annotated or incomplete pathways).

Functions to accomplish these tasks are provided in the script get_panEFMs.py.

1) Load the *Bacteroides* pan-reactome model and generate random environments (as we did previously). We use five randomly sampled environments as illustration, though in practice a larger number is recommended (e.g., 1000).

```python
from scripts.get_panEFMs import *

model     = cobra.io.read_sbml_model(model_path)
exchanges = get_exchange_metabolites(model)
```

```python
envBall = gen_environment_ball(
    exchanges,
    anaerobic       = True,
    fixed_reactions = {'EX_h2o(e)': 100},
    size            = 5,
    total_flux      = 100,
    seed            = 666
)
```

2) Sample pan-EFMs for each environment. We demonstrate with 10 samples per environment (use a larger number, e.g., 1000, in practice). To sample a pan-EFM, we randomly shuffle the list of reactions and attempt to remove them one by one. If removing a reaction causes the objective value to fall below a defined threshold, the reaction is marked as essential and kept; otherwise, it is removed. The result is a binary vector indicating which reactions are essential for that specific minimal growth-supporting configuration. Repeating this process produces a set of pan-EFMs for each environment.

```python
pan_efms = {}
for i in envBall:
    print(f"Environment Simulation: {i}")
    pan_efms[i] = get_panEFM_dist(model_path, reactions, envBall[i], max_it=10)
```

3) Summarize and visualize panEFMs across environments. To explore how reaction essentiality varies across environments, we calculate the frequency with which each reaction appears in the sampled pan-EFMs for each condition. We visualize the resulting matrix as a heatmap, with reactions sorted from left to right by their average frequency across environments. Reactions with high frequency are consistently essential across all environments. Reactions with intermediate frequency are conditionally essential—they are required in some environments but not others. Low-frequency reactions are mostly dispensable and rarely needed for growth.

```python
frequency_df = pd.DataFrame({
    env: np.mean(pan_efms[env], axis=0)
    for env in pan_efms
    },
    index=reactions
).T

results_folder = root_dir / 'results' / 'pan_efms'
os.makedirs(results_folder, exist_ok=True)

output_path = results_folder / 'pan_efms_dist.png'
plot_reaction_freq_heatmap(
    frequency_df, output_path=output_path, figsize=(12, 8), cmap="coolwarm"
)
```

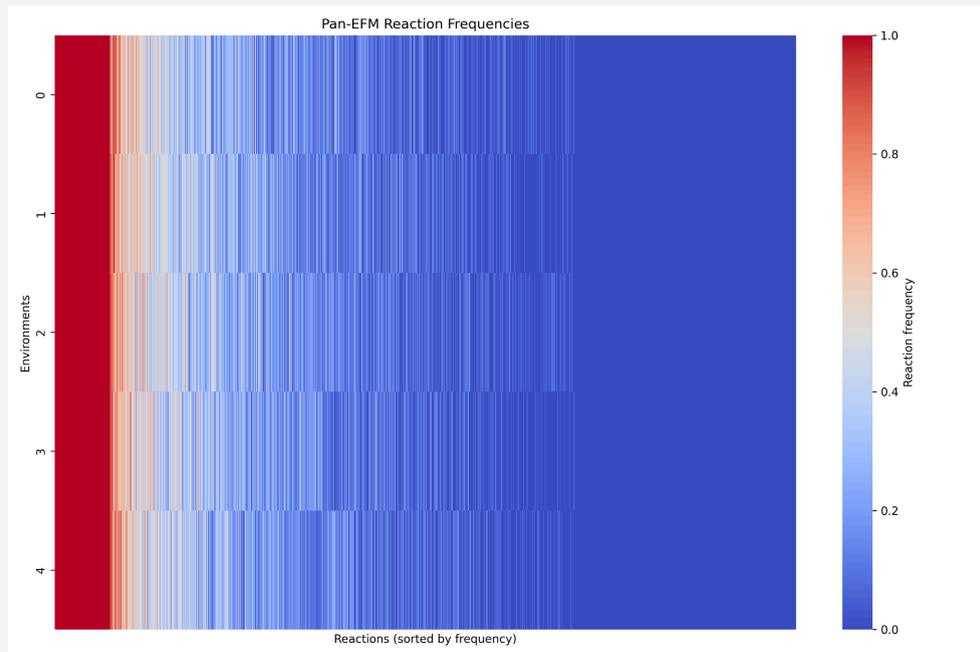

**Fig. 6: Reaction frequencies in pan-EFMs.** Reactions with zero frequency (dark blue) were always dispensable, while reactions with a frequency of one (dark red) are essential independently of the environment. Reactions with intermediate frequency are conditionally essential.

In the same notion, recently, Øyås et al. introduced the **minimal pathways (MPs)** [29]; minimal subsets of reactions from a metabolic **sub-**network. Thus, minimal pathways need to include at least one functional requirement, i.e., a constraint requiring a metabolic process to be performed at least to some extent, and only reactions with non-zero flux. This way, if one could enumerate all the different MPs of a model, they could distinguish essential reactions from those with relative functional importance and others that never actually contribute. Combining MPs enumeration with flux sampling provides insight in predicting growth effects of gene knockouts (species level), and uncovering metabolic interactions (community level). For a thorough example on how to perform this, the reader may follow the steps discussed in the uni- and multicellular_analysis_ove notebooks of this GitLab repository[6].

## Sampling the environment in a community context

As a final example of sampling genome-scale metabolic models, we shift our focus to microbial communities. In many microbiome studies, we obtain taxonomic compositions of samples but lack precise information about absolute abundances. A natural question is: can we use genome-scale metabolic models to explore which steady-state metabolite environments are consistent with an observed species composition?

This is a compelling application of sampling, because there are typically many possible environments that can give rise to a given community structure. By sampling these environments, we can explore the distribution of metabolite profiles that support a desired species composition, and identify which metabolites tend to be over- or under-expressed across conditions.

---

[6] https://gitlab.com/YlvaKaW/exchange-enumeration

It is important to remember that genome-scale models operate under steady-state assumptions, and realistic environmental constraints are difficult to define in a community context. In our view, one cannot assume that dietary profiles or even bulk metabolite concentrations accurately represent the true constraints experienced by individual microbial populations. To address this, we use sampling to flexibly explore the space of feasible environments, searching for those that produce steady-state biomass outputs most correlated with a target community composition. Each species is modeled independently, constrained by the same environment, and allowed to optimize its own growth.

To illustrate this concept, we use the MAMBO algorithm to sample metabolite environments that support specific species compositions. We use three models representing different functional guilds: *Blautia hydrogenotrophica* (acetogen), *Bacteroides thetaiotaomicron* (sugar fermenter), and *Roseburia intestinalis* (butyrate producer). As an illustrative example, we can test different target compositions in which different species dominate, e.g.: [10, 1, 0.1], [1, 10, 0.1], [1, 0.1, 10]. Below, we illustrate the first one.

For each case, we can sample environments using MAMBO [30] and plot the metabolite profiles of the top 1% of samples with the highest correlation to the desired species composition.

Functions to accomplish this task are provided in the script mambo.py.

1) Import the script and load the three models:

```python
from mambo import *

# Load species models
acetogen = cobra.io.read_sbml_model(
    agora_models_path / 'Blautia_hydrogenotrophica_DSM_10507.xml'
)
sugar_fermenter = cobra.io.read_sbml_model(
    agora_models_path / 'Bacteroides_thetaiotaomicron_VPI_5482.xml'
)
butyrate_producer = cobra.io.read_sbml_model(
    agora_models_path / 'Roseburia_intestinalis_L1_82.xml'
)
```

2) Extract exchange reactions and define a medium dictionary as a starting point. Here we set all metabolite concentrations to one.

```python
exchanges = get_exchange_metabolites(acetogen)
exchanges = exchanges.union(get_exchange_metabolites(sugar_fermenter))
exchanges = exchanges.union(get_exchange_metabolites(butyrate_producer))

media = {i: 1 for i in exchanges}
```

3) Apply the starting media to the three models, optimize them and store the models in a list. The order of this list is critical and should match with the measured composition. We also define a composition vector. In the case below, the acetogen wins, changing the vector for the other scenarios.

```
acetogen.optimize()
sugar_fermenter.optimize()
butyrate_producer.optimize()

# Order is critical
modelList = [acetogen, sugar_fermenter, butyrate_producer]

# Composition Vector
composition = np.array([10, 1, 0.1])

medias    = [np.array(list(media.values()))]
solutions = [current_solution(modelList, media)]
```

4) Run the MAMBO optimization. We run for 5,000 iterations (10,000 for the case where the butyrate producer wins); by increasing this number, we reach more complex samples.

```
for i in range(5000): #should be much larger
    solution, media = MCMC(media, modelList, composition, delta = 1)

    if (i>10):  # Should be much larger
        medias.append(np.array(list(media.values())))
        solutions.append(solution)
```

5) We normalize the sampled media and biomass vectors, then compute the correlation between predicted and target composition.

```
medias  = np.array(medias)
mediasM = medias.copy()
medias  = medias.T

maxMedias = np.max(mediasM, axis=1)
mediasM   = np.array(
    [mediasM[i]/maxMedias[i] for i in range(len(maxMedias))]
).T

mSols     = solutions.copy()
solutions = np.array(solutions).T
maxSolutions = np.max(mSols, axis=1)
mSols     = np.array(
    [mSols[i]/maxSolutions[i] for i in range(len(maxSolutions))]
).T
cor = np.array(
    [sts.pearsonr(i, composition)[0] for i in solutions.T]
)
```

6) Compute the median environment from the top 1% samples. This tells us which metabolites are consistently enriched in environments that support the desired biomass structure.

```
avM = np.median(medias.T[cor>max(cor)*0.99], axis=0)
avM = (avM/max(avM))*10
m   = {list(media.keys())[i]: avM[i] for i in range(len(avM))}

print(f"composition was {composition} \t MAMBO solution was: {current_solution(modelList, m)}")
```

7) Plot the results, sorted by the correlations between the steady state biomass prediction and the measured composition.

```
sorter     = np.argsort(cor)
result     = medias.T[sorter].T
cor_sorted = cor[sorter]

results_folder = root_dir / 'results' / 'mambo'
os.makedirs(results_folder, exist_ok=True)

output_path = results_folder / 'mambo_acetogenWins.png'
plot_mambo_results(result, cor_sorted, output_path=output_path)
```

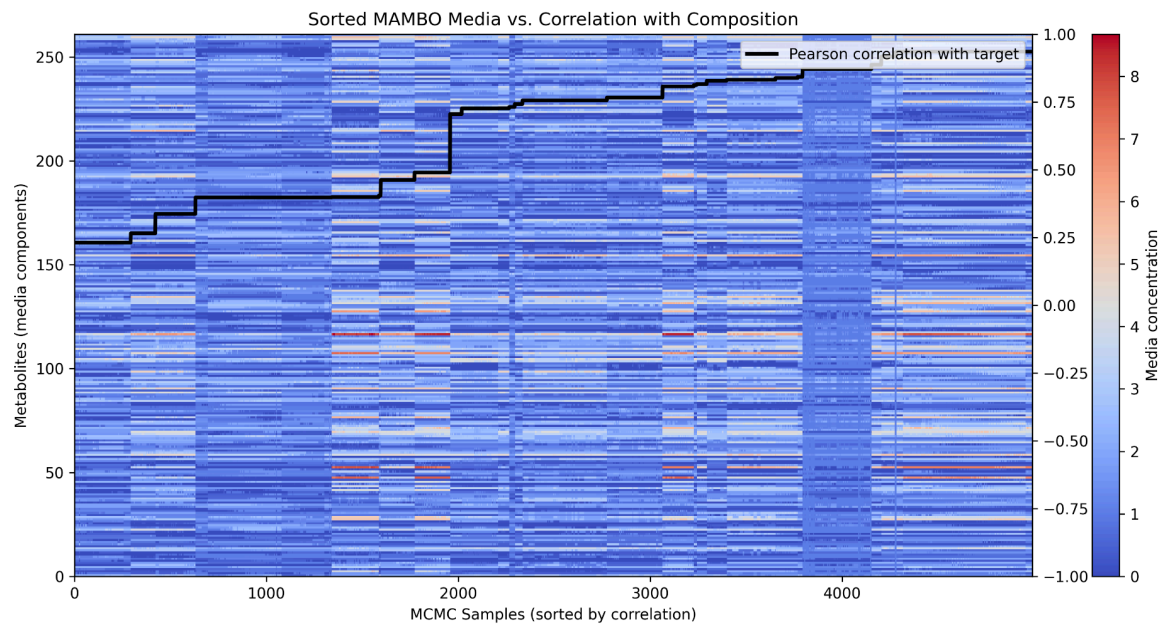

**Fig. 7: Concentration of extracellular metabolites that support a defined steady-state biomass composition.** Metabolite compositions were inferred by the MAMBO algorithm and are sorted by their correlation between the steady-state biomass fluxes in the in silico environment and the relative abundance of the species (usually measured by sequencing). The black line represents the correlation values. In this case, the acetogen, Bh, it the one that "wins" in relative abundance